\documentclass[a4paper,12pt]{article}

\usepackage[utf8]{inputenc}
\usepackage[T1]{fontenc}
\usepackage{float}
\usepackage{amsmath,amssymb}
\usepackage{mathrsfs}
\usepackage{theorem}
\usepackage{graphicx}
\usepackage{color}
\usepackage{wasysym}
\usepackage{hyperref} 
\usepackage{verbatim}
\usepackage{datetime2}
\definecolor{blue}{rgb}{0,0,1}
\definecolor{green}{rgb}{0,0.65,0.5}
\definecolor{verde}{rgb}{0.,.5,0.4}
\definecolor{marron}{rgb}{0.7,0.2,0.1}
\definecolor{red}{rgb}{1,0,0}
\definecolor{vio}{rgb}{1,0,1}
\definecolor{ama}{rgb}{1,1,0}

{\theorembodyfont{\sffamily} \newtheorem{teorema}{Theorem}[section]}
{\theorembodyfont{\sffamily} \newtheorem{corolario}{Corollary}[section]}
{\theorembodyfont{\sffamily} }
{\theorembodyfont{\sffamily} \newtheorem{definicion}{Definition}[section]}
{\theorembodyfont{\sffamily} }
\begin{document}

\title{\bf 
	Geometry of transient gravitational waves and
	estimation of efficiencies of different detector configurations
% \\
%  {\sf\small  (de: rel-part-pard-1-2.tex)}
}

\author{
Osvaldo M. Moreschi\thanks{Email: o.moreschi@unc.edu.ar}%\thanks{Member of CONICET.}
\\
{\rm \small Facultad de Matemática Astronomía, Física y Computación (FaMAF)} \\
{\rm \small Universidad Nacional de C\'{o}rdoba}\\
{\rm \small Instituto de F\'\i{}sica Enrique Gaviola (IFEG)}\\
{\rm \small Ciudad Universitaria, (5000) C\'{o}rdoba, Argentina}
}

\date{April 27, 2026}
\maketitle

%\date{\DTMnow} % con \usepackage{datetime2} % comentado para arxiv

\vspace{3mm}
%\keywords{Gravitational waves, Gravitational wave astronomy, Astronomy data analysis}
{\sf keywords:} Gravitational waves, Gravitational-wave astronomy, Gravitational-wave observatories

\begin{abstract}

This work introduces a geometrical method for analyzing transient gravitational waves 
recorded at interferometric observatories. 
This approach is intended to aid in 
assessing the performance and sensitivity of next-generation detector configurations, 
such as Cosmic Explorer, Einstein Telescope, and 
the South American Gravitational-wave Observatory.

\end{abstract}

{\sf \footnotesize %\scriptsize
	\tableofcontents
}

\section{Introduction}

We define transient gravitational waves (TGW) as short-lived signals with durations 
ranging from several seconds to approximately one minute. 
This term captures the various nomenclature used in the literature, such as 
'transients' or 'bursts', to describe compact binary coalescences. 
To date, all confirmed gravitational-wave detections are classified as TGW.

Currently, a global network of large-scale interferometric observatories is operational, 
having completed several successful observing runs. 
This network includes the 
\href{https://ligo.org}{LIGO} ,
\href{https://www.virgo-gw.eu}{Virgo}, and 
\href{https://gwcenter.icrr.u-tokyo.ac.jp/en/}{KAGRA}
detectors. 
During the recently concluded fourth observing run (O4), the collaboration 
announced more than 
\href{https://www.virgo-gw.eu/news/record-detection-of-200-gravitational-waves-in-the-current-run-of-ligo-virgo-and-kagra/}{200 candidate events}, 
significantly expanding 
the TGW catalog.

Proposed third-generation (3G) gravitational-wave (GW) observatories, such as the 
\href{https://www.et-gw.eu}{Einstein Telescope}\cite{Branchesi:2023mws, ET:2025xjr}
and 
\href{https://cosmicexplorer.org}{Cosmic Explorer},
provide strong impetus for ongoing studies into their scientific potential and 
optimal configurations\cite{Freise:2008dk, Freise:2010viw, Branchesi:2023mws}. 
Recently, at the First International Latin American Conference on Gravitational Waves
(\href{agenda.infn.it/event/45592/}{1st LAGW}),
the initiative for a South American Gravitational-wave Observatory 
(\href{https://www.ictp-saifr.org/wp-content/uploads/2020/07/09-Odylio-D.-Aguiar.pdf}{SAGO})
was revitalized. 
This renewed interest motivates our investigation into the sensitivity and performance 
of various detector designs situated at southern latitudes.

While several other gravitational-wave observatories are currently 
in various stages of development or employ alternative detection techniques, 
this theoretical study focuses specifically on the geometric 
configurations of the third-generation detectors mentioned above.

In this article, we present a novel geometrical framework for analysis of TGWs. 
This framework introduces new theoretical concepts and establishes fundamental theorems 
that facilitate the evaluation of sensitivity and efficiency across diverse detector geometries. 
We provide initial results for proposed 3G configurations.
A comprehensive analysis incorporating specific detector dimensions, 
precise geolocations, and cross-observatory comparisons is reserved for future work.

Beyond theoretical design, the techniques presented here offer a new  perspective for
examining recorded GW events. 
A detailed application of this approach to existing catalog data 
is currently being prepared for future publication.

The remainder of this article is organized as follows. Section \ref{sec:geometry}, 
introduces several new concepts for gravitational-wave detection and derive fundamental 
results associated with these frameworks. 
Section \ref{sec:applications} applies these theoretical findings to various 
geometric configurations currently under consideration for Cosmic Explorer, 
Einstein Telescope, and SAGO projects. 
Section \ref{sec:final}
recapitulates our main findings and discuss their implications.

%%%%%%%%%%%%%%%%%%%%%%%%%%%%%%%%%%%%%%%%%%%%%%%%%%%%%%%

%\linea
\section{Intrinsic geometry in GW detections}\label{sec:geometry}

\subsection{Basic algebra in a detection of a GW}

The framework presented here is designed for application to existing observatories, 
such as LIGO Hanford ($H$), LIGO Livingston ($L$), Virgo ($V$), and KAGRA ($K$), 
as well as proposed next-generation projects including 
Cosmic Explorer ($C$), Einstein Telescope ($E$), and SAGO ($S$).

When observatory $X$ records a TGW, the measured strain, 
over the time interval of interest $\Delta t$
is represented as the sum of noise and the GW signal.
The noise is assumed to be of stochastic nature.
The signal instead is assumed to be expressed solely through the modulation
of the pattern functions $F_{+X}$ and $F_{\times X}$ over
the GW polarization modes (PM) $s_+$ and $s_\times$.
More concretely, the strain is expressed as:
\begin{equation}\label{eq:vX}
\begin{split}
v_X(t + \tau_X) &= n_X(t + \tau_X) + s_X(t + \tau_X) \\
&= n_X(t + \tau_X) + 
F_{+X}(\theta_X,\phi_X,\psi_X) s_+(t) \\
&\;\;\; +
F_{\times X}(\theta_X,\phi_X,\psi_X) s_\times(t)
,
\end{split}
\end{equation}
where $X$ denotes the observatory identifier and
$\tau_X$ is the time delay of detector $X$ with respect to
the chosen reference observatory. 
The coordinates
$(\theta_X,\phi_X,\psi_X)$ are the angular coordinates with respect
to detector $X$ at the reference event time $t_e$.
Specifically, the first two angular coordinates determine
the direction of the source, while the third is the angle of the
GW frame, sometimes called polarization angle or, more precisely, the polarization frame angle;
$t$ is the time variable.
The measured strain is denoted by $v$, the noise by $n$, and the GW signal by $s$,
which is decomposed in the PMs $s_+$ and $s_\times$.
The detector pattern functions\cite{Poisson2014} $F_+$ and $F_\times$ are  
recalled in Appendix \ref{app:pattern} ,
where the expressions 
account for an arbitrary opening angle $\chi$
between the interferometer arms that may differ from $\pi/2$.
When $\Delta t$ is very small, the observatories make a negligible change of orientation
during the recording of the GW signal, due to Earth rotation,
and therefore the angular coordinates denoting
the position of the source  remain nearly constant and we can safely use the orientation angles
as determined at the reference event time $t_e$.

Conversely, if  $\Delta t$  exceeds approximately four minutes, then one should incorporate the
time dependence $t + \tau_X$ in the arguments of the pattern functions.
However, as the primary objective of this work is to establish convenient concepts 
for evaluating the sensitivity of diverse observatory configurations, 
we restrict our analysis to the regime of short-duration (transient) GW signals. 
Nonetheless, the geometrical framework developed herein can be naturally extended 
to signals persisting over longer durations by incorporating 
the time-varying orientation of the detector frames.

\subsection{Geometrical definitions for a GW event}

In the previous expressions, instead of treating $s_+(t)$ and $s_\times(t)$ as continuous functions,
one can consider the multidimensional vectors $\vec{s}_+$ and $\vec{s}_\times$
defined as the corresponding discretized signals $s_+$ and $s_\times$ of length $N= f\!s \, \Delta t$,
where $f\!s$ is the sampling rate. For current observatories,
$f\!s = 16384$ samples per second.
While the Virgo detector natively acquires data at 
20000 samples per second(\href{https://dcc.ligo.org/public/0027/T050264/002/T050264-07.pdf}{technical report T050264-07}),
publicly released strains are resampled to the standard rate of 16384. 
Using this discrete geometrical representation of the signal contained in the recorded strain,
one can express these two vectors as $\vec{s}_+ = \varsigma_+ \hat{s}_+$ and 
$\vec{s}_\times = \varsigma_\times  \hat{s}_\times$;
where the $\hat{s}_+$ and $\hat{s}_\times$ are unit vectors in the 2-dimensional subspace spanned by
$\vec{s}_+$ and $\vec{s}_\times$, 
and therefore $\varsigma_+$ and $\varsigma_\times$ are the respective magnitudes;
that carry the detail time-dependent information of the polarization modes.

Then, we define:
\begin{definicion}
The `plane of the GW event' is the 2-dimensional space
generated by the orthonormal vectors $\hat{s}_+$ and $\hat{s}_\times$.
\end{definicion}
Note that by definition the plane of the GW event has a natural
Euclidean geometry; which allows to measure angles and norms.

In this plane, associated with each event, we also define: 
\begin{definicion}
For each detector $X$, the `detector vector' is the linear combination given by
\begin{equation}\label{eq:vec_sX}
	\vec{s}_X = 
	F_{+X}(\theta_X,\phi_X,\psi_X) \hat{s}_+ 
	+
	F_{\times X}(\theta_X,\phi_X,\psi_X) \hat{s}_\times
	.
\end{equation}
\end{definicion}
Using this notation, $\hat{s}_+$ and $\hat{s}_\times$  inherit the 
transformation properties of the PMs + and $\times$
under changes in the $\psi$ angle, recalled in Appendix \ref{app:transfpm}.

Given the \emph{detector vectors} $\vec{s}_{X1}$ and $\vec{s}_{X2}$, we define:
\begin{definicion}\label{def:scalar}
The scalar product of the two \emph{detector vectors} $\vec{s}_{X1}$ and $\vec{s}_{X2}$
is given by
\begin{equation}\label{eq:scalar_X1X2}
	< \vec{s}_{X1},\vec{s}_{X2}> = F_{+X1} F_{+X2} + F_{\times X1} F_{\times X2}
	,
\end{equation}
where we use $<.,.>$ to denote scalar products.
\end{definicion}
This is the natural scalar product in the 2-dimensional
space spanned by the orthonormal vectors $\hat{s}_+$ and $\hat{s}_\times$.

We also define:
\begin{definicion}\label{def:corr}
The correlation $\rho_{X1,X2}$ is given by
\begin{equation}\label{eq:corr_X1X2}
\rho_{X1,X2}  = \frac{ < \vec{s}_{X1},\vec{s}_{X2}>  }
{\sqrt{< \vec{s}_{X1},\vec{s}_{X1}>} \sqrt{< \vec{s}_{X2},\vec{s}_{X2}>}}
,
\end{equation}
where it can be seen that the correlation
coincides with the cosine of the angle between the two \emph{detector vectors}.
\end{definicion}
The relevance of the correlation between two \emph{detector vectors}
is that it measures the ability to distinguish two PMs
in the presence of two signals, of the same GW, at corresponding observatories.
For example a correlation of 1 or -1 would imply that the two observatories
have recorded the same (up to sign) PM component.
Consequently, correlations close to zero 
represent the optimal situation for
distinguishing the two PM components.

If one restricts attention to the 2-dimensional \emph{plane of the GW event},
then the exterior product of two \emph{detector vectors} yields
a pseudoscalar.
That is, we define:
\begin{definicion}\label{def:wedge}
The exterior product $w_{X1,X2}$ between the two
\emph{detector vectors} $\vec{s}_{X1}$ and $\vec{s}_{X2}$
 is given by
\begin{equation}\label{eq:w_X1X2}
w_{X1,X2}  = F_{+X1} F_{\times X2} -  F_{+X2} F_{\times X1}
.
\end{equation}
\end{definicion}
This exterior product plays a crucial role in the algebra involved in the
reconstruction\cite{Moreschi:2025fxg} of the GW polarization modes coming from observations in
two observatories; as it coincides with the determinant of the
matrix relating PMs to observed signals.

These definitions can be applied to study the properties of different
gravitational-wave observatory configurations
using their specific geolocations.

\subsection{Theoretical results from the new concepts}
The advantage of introducing these geometrical interpretation of recorded GWs
is that one can readily derive useful properties for GW data analysis.
For example, consider the following theorem.

\begin{teorema}\label{teo:vec_inva}
The \emph{detector vector} $\vec{s}_X$ is polarization-frame invariant;
that is, it is independent of the angle $\psi$.
In other words, it is a genuine vector in the \emph{plane of the GW event}.
\end{teorema}
The proof follows from the transformation properties of the
basis vectors $\hat{s}_+$ and $\hat{s}_\times$ and the components
$F_{+X}$ and $F_{\times X}$, where the latter are discussed in Appendix
\ref{app:transfpattfunc} .
More concretely, it is easy to see that while the vector frame
$\big(\begin{smallmatrix} \hat{s}_+ , \, \hat{s}_\times \end{smallmatrix}\big)$
transforms with a rotation matrix $R$,
the vector components
$\big(\begin{smallmatrix} F_+ , \, F_\times \end{smallmatrix}\big)$
transforms via the inverse rotation matrix $R^{-1}$.

This result can also be understood in terms of the observed signal $s_X$:
\begin{corolario}\label{cor:indepen}
The recorded signal $s_X$ is independent of the chosen GW frame
used in the polarization mode decomposition.
\end{corolario}
This follows from the fact that 
$s_X = \big(\begin{smallmatrix} \varsigma_+ \hat{s}_+ \,\; \varsigma_\times \hat{s}_\times \end{smallmatrix}\big)
\big(\begin{smallmatrix} F_{+X} \\ F_{\times X} \end{smallmatrix}\big)$.
One could also equivalently argue that $s_X$ is an observable that does not involve
any choice of GW frame.

Another useful result is the following theorem that relates the detector vectors
of two observatories with the same origin and coplanar geometry.
\begin{teorema}\label{teo:2vec}
Let there be two detectors $X1$ and $X2$ 
of the same dimensions and geometry,
sharing the same origin
and lying in the same plane, and with the angle bisector of $X2$ 
rotated by $\Delta \phi$ from the angle bisector of $X1$.
Let the \emph{detector vector} $\vec{s}_{X1}$ be determined by angular coordinates
$(\theta,\phi,\psi)$.
Then, the \emph{detector vector} $\vec{s}_{X2}$ is located on the ellipse
of semi-major axis $\frac{1}{2}\big(1+\cos(\theta)^2 \big) \sin(\chi)$ 
and semi-minor axis $|\cos(\theta)  \sin(\chi)|$, 
where  $\chi$ is the opening angle between the observatory arms.
And their scalar product is:
\begin{equation}\label{eq:s1dots2_1}
\begin{split}
< \vec{s}_{X1} ,	\vec{s}_{X2}> 
=&
\Big(\frac{1}{2}( 1 + \cos(\theta)^2)  \sin(\chi) \Big)^2
\cos(2 \phi) \cos(2 \phi - 2 \Delta \phi) \\
&+
\Big( \cos(\theta)  \sin(\chi) \Big)^2
\sin(2 \phi)\sin(2 \phi - 2 \Delta \phi)
.
\end{split}
\end{equation}

\end{teorema}
The elements of the proof are presented in Appendix \ref{app:transfphi} ;
where one can see that the angle of the direction of $\vec{s}_{X2}$
with respect to $\vec{s}_{X1}$ is not in general $2\Delta \phi$
as one might have guessed.

Note that the ellipse degenerates to a line segment in special case where $\theta = \frac{\pi}{2}$.
In other words, when the source of the GW is located in the equatorial plane
of the observatories, only one polarization component can be detected by
this geometric arrangement.
At the other extreme, the ellipse becomes a circle at $\theta = 0$
and $\theta=\pi$.

The notion of polarization frame angle and its use have been the source of
misleading and erroneous statements in the literature.
For this reason it is useful to state explicitly a result
that clarifies several discussions:
\begin{teorema}\label{teo:2vec_samepsi}
	Let two detectors $X1$ and $X2$ 
	share the same origin and lie in the same plane.
	Then, $\psi_{X1}=\psi_{X2}$; where $\psi_X$ is the polarization frame angle appearing in equation\eqref{eq:vX}.
\end{teorema}
This follows directly from the very definition of polarization frame angle\,\cite{Poisson2014}.

The following corollaries are noteworthy: 
\begin{corolario}\label{cor:scalar_indepen}
	The scalar product $<\!\vec{s}_{X1},\vec{s}_{X2}\!>$
	is independent of the chosen  polarization frame. 
\end{corolario}

\begin{corolario}\label{cor:corr_indepen}
	The correlation $\rho_{X1,X2}$
	is independent of the chosen  polarization frame. 
\end{corolario}
\begin{corolario}\label{cor:w_indepen}
	The exterior product $w_{X1,X2}$
	is independent of the chosen  polarization frame. 
\end{corolario}

These results prove particularly valuable when analyzing observatories with diverse detector configurations.

\section{Applications to different observatory configurations}\label{sec:applications}

\subsection{The case of two L-shaped detectors at $\frac{\pi}{4}$ angle}\label{sec:2L}

When applying the previous results to two L-shaped observatories
with the same physical dimensions, sharing a common origin and lying in the same plane,
and at an  angle $\Delta \phi = \pm \frac{\pi}{4}$, one deduces (See appendix \ref{app:transfphi}.) 
that the 
scalar product between the detector vector $\vec{s}_{X1}$
and $\vec{s}_{X2}$ is
\begin{equation}\label{eq:s1dots2_pis2_1}
	\begin{split}
		< \vec{s}_{X1} ,	\vec{s}_{X2}>
		=&
		\pm \frac{1}{8} \sin(4 \phi)  \sin(\chi)^2 \sin(\theta)^4
		.
	\end{split}
\end{equation}
This function has a maximum amplitude of 0.125;
where the $\theta$ contribution emphasizes the equator
and vanishes at the poles,
while the $\phi$ contribution gives an oscillating modulation. 
This indicates that the scalar product has relatively low amplitude across the sky, 
representing favorable conditions for distinguishing polarization mode components.
Detailed analysis of these effects for specific observatory projects 
will be presented in a separate article.

\begin{figure}[h]
\centering
\includegraphics[clip,width=0.46\textwidth]{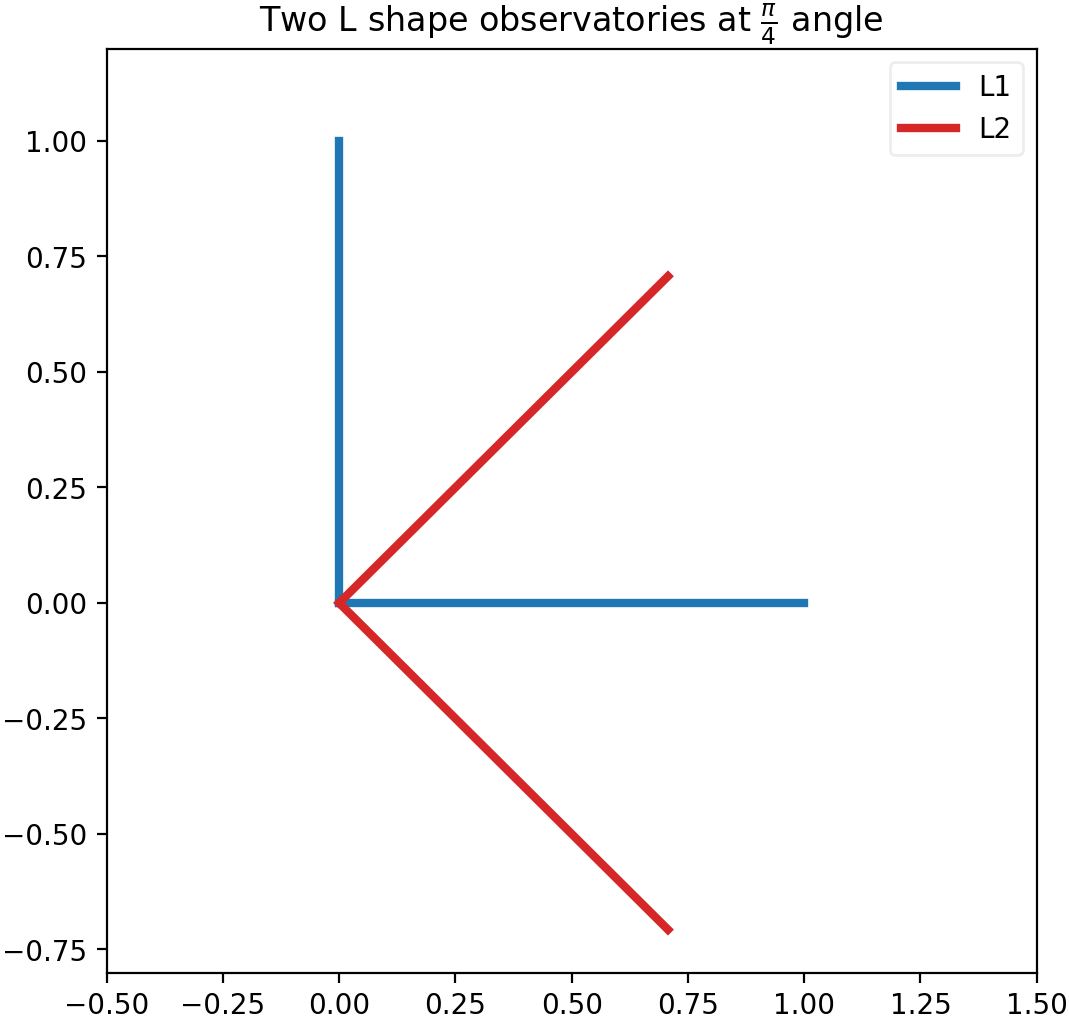}
\caption{Sketch of two L shape GW detectors with middle line 
	at relative $\frac{\pi}{4}$ angle.
}\label{fig:2L}
\end{figure}

The analysis of signals from two co-located gravitational-wave detectors 
also offers the advantage of enabling filtering techniques for common seismic noise, 
potentially improving the detectability of GW signals at low frequencies.

\subsection{The case of three detectors in a triangular configuration at $\frac{2\pi}{3}$ angle}\label{sec:3T}

The equilateral triangular configuration for GW observatories offers 
the natural benefit of housing three detectors at essentially the same location. 
In this case, while any pair of detectors does not share exactly the same origin, 
the separation between them is only one arm length, which is negligible compared 
to Earth's dimensions. Consequently, the previous results can be safely applied.

\begin{figure}[h]
\centering
\includegraphics[clip,width=0.46\textwidth]{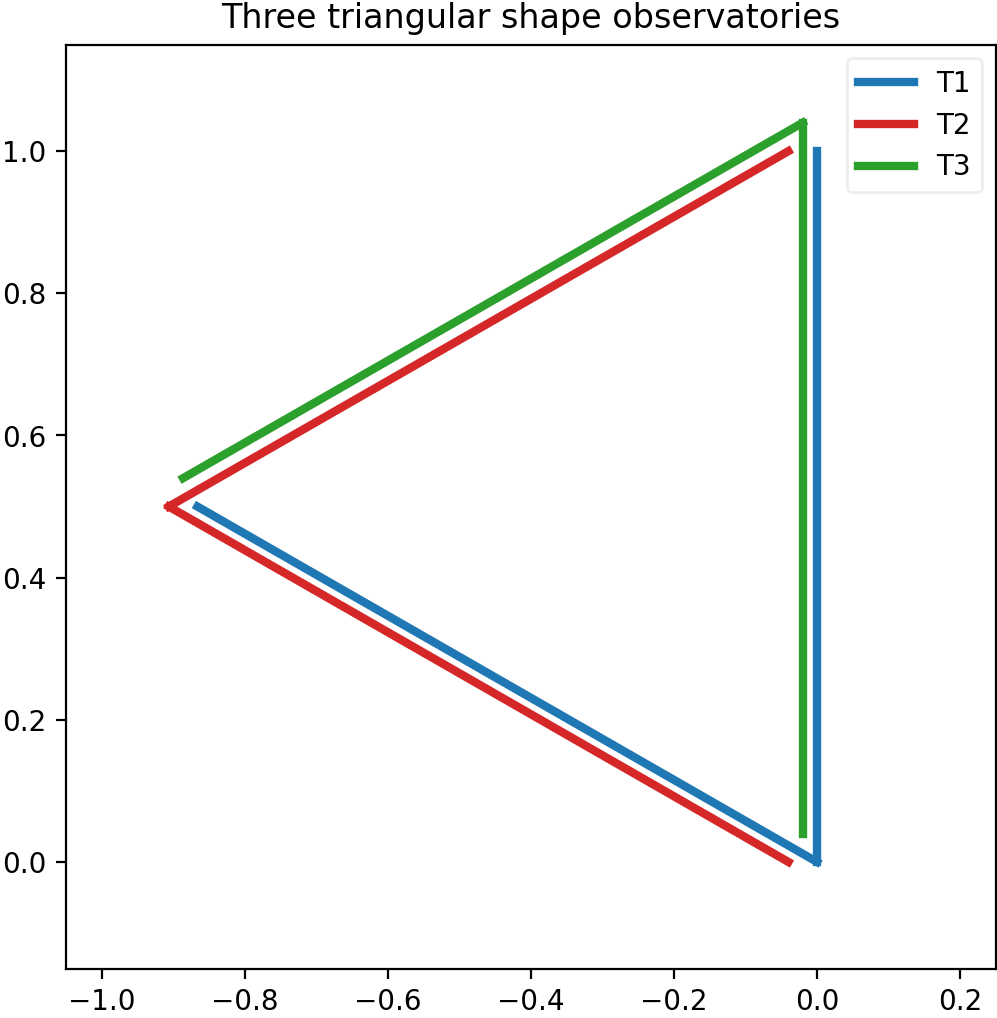}
\caption{Sketch of three GW detectors in an equilateral triangular shape array.
}\label{fig:3T}
\end{figure}

One can then consider the relative angle between two observatories to be $\frac{2\pi}{3}$.
As noticed previously, we cannot assume that in general
the second \emph{detector vector} $\vec{s}_{X2}$
is oriented at an angle of $\frac{4\pi}{3}$ with respect to  $\vec{s}_{X1}$. %;
%or equivalently that $\vec{s}_{X2}$ would be at an angle $\frac{-2\pi}{3}$
%with respect to  $\vec{s}_{X1}$.
Nevertheless, one can calculate the sum of the three \emph{detector vectors}.
The simple algebra presented in Appendix \ref{app:triangular} shows that 
the sum of the strains recorded at the three detectors 
in an equilateral triangular configuration yields the null stream:
\begin{equation}\label{eq:sumxsig}
		s_{X1} + s_{X2} + s_{X3} = 0
	.
\end{equation}
That is, this simple algebraic operation cancels the GW signal in principle. 
While calibration uncertainties may complicate this theoretical result in practice, 
conversely, this null-stream property can be exploited for fine-tuning 
relative calibration between detectors.

Since gravitational waves have two degrees of freedom encoded in the two polarization modes, 
and each observatory detects only one linear combination of these modes, 
one might consider, when analyzing signals from three observatories, 
finding the particular linear combination of the three strains that cancels the GW signal 
as an intermediate analysis step. 
In fact, this approach was suggested in \cite{Gursel:1989} as a method to solve 
the inverse problem for GW detection, namely, to determine the sky location 
and polarization mode components using three detectors. 
However, it was shown in \cite{Moreschi:2025fxg} that for generic 
chirp-like signals, the inverse problem can actually be solved using 
data from only two observatories.

In principle, with three observatories at different locations, one can use two of them 
to determine the sky location and polarization modes, then test consistency with 
the signal recorded at the third observatory. 
However, in the general case, one must confront the difficulties of time delays 
between observations and the sky-location dependence of the measurements. 
These difficulties vanish when the three observatories share the same geographical location. 
As shown for the equilateral triangular configuration, the null stream is satisfied 
for all sky locations and involves no time delays. 
In other words, the triangular design is highly advantageous for analyzing %\cite{Chatterji:2006nh} 
detected gravitational waves, since it provides access to relations among the signals that 
are completely model-independent\cite{Freise:2008dk} and therefore require no 
assumptions about the signal beyond its polarization properties as a gravitational wave.

The null stream can be used to eliminate transient detector 
glitches\cite{Goncharov:2022dgl}, subtract 
loud non-Gaussian noise from the data, and resolve calibration 
issues\cite{Branchesi:2023mws}.%\cite{Chatterji:2006nh,Branchesi:2023mws}.

Similarly, 
this implies that in search mechanisms\cite{Moreschi:2025fxg} for the \emph{detector vector}, 
one could improve the techniques by exploiting the known relative angles among the three \emph{detector vectors}.

Having three detectors in close proximity 
enables studies of their correlated noise\cite{Janssens:2022cty} and, 
as mentioned previously, allows for the design of improved filtering techniques 
to eliminate common seismic noise, particularly at low frequencies.

It should be noted that having a $\frac{\pi}{3}$ angle between the arms
of one observatory diminish the amplitude of the pattern functions by
a factor of approximately $0.866$.
But this is compensated by the advantage of having three signals
in the triangular array.

\subsection{The case of a tristar configuration}

We propose to include in the discussion the tristar configuration, which 
consists of three detectors sharing a common origin with arms at $\frac{2\pi}{3}$ angle,
such that each of the three legs contains two arms from adjacent detectors. 
The sensitivity reduction factor is identical to that of the triangular configuration, 
since $\sin(\frac{\pi}{3}) = \sin(\frac{2\pi}{3})$.
The immediate advantage of the tristar configuration over the triangular one 
is that the central units for the three detectors are in close proximity, 
rather than being separated by approximately 10 km (one arm length).
This configuration offers significant operational advantages: 
simplified construction logistics, centralized maintenance, 
and elimination of inter-site timing synchronization challenges.

\begin{figure}[H]
\centering
\includegraphics[clip,width=0.46\textwidth]{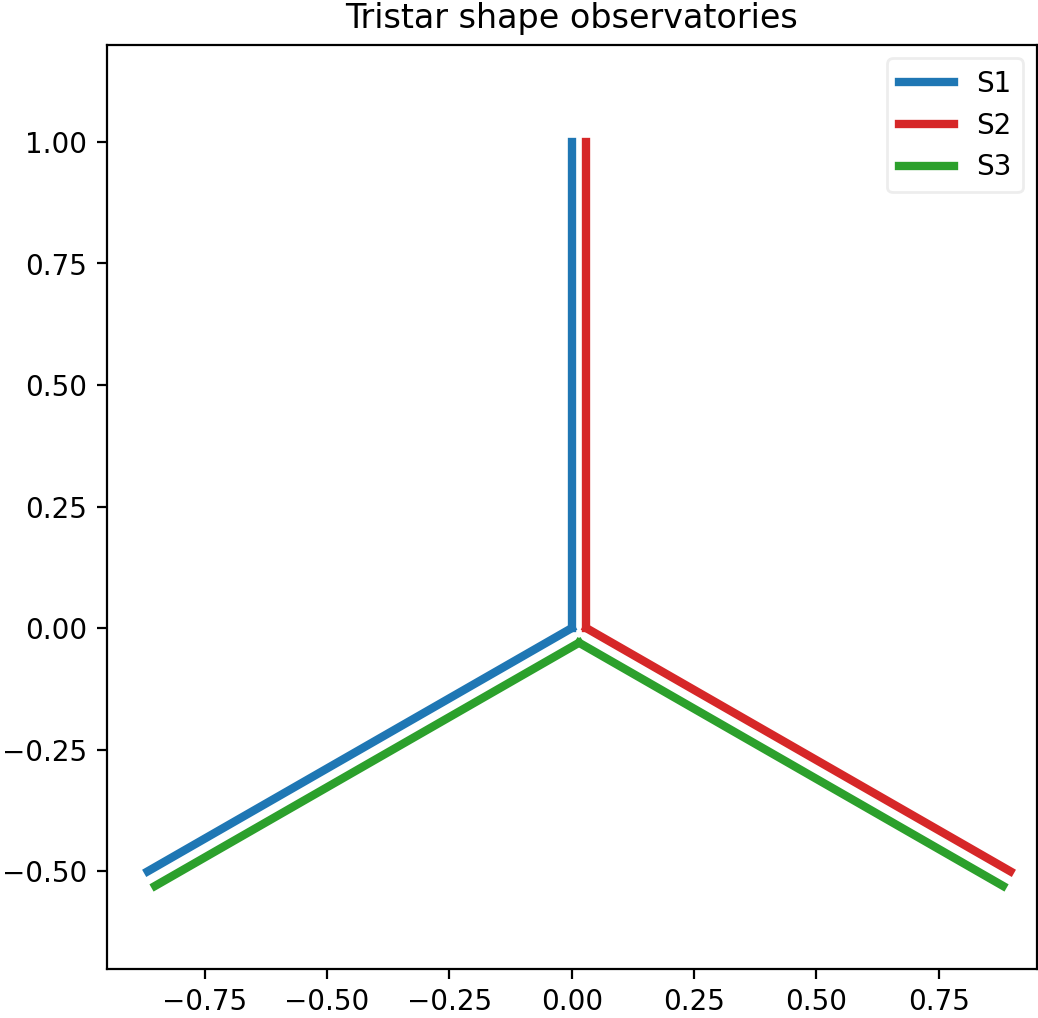}
\caption{Sketch of three GW detectors in a tristar configuration array.
}\label{fig:3S}
\end{figure}
Since the relative angle between detectors is also  $\frac{2\pi}{3}$,
all the discussion of the triangular configuration applies here as well, 
including the null stream formula. 

Having multiple detectors at the same site enables, as previously mentioned, 
the development of enhanced techniques for filtering, source localization, 
and polarization mode reconstruction.

\subsection{Angular sensitivity of present an planned detectors}

For the study of the angular sensitivity of different observatory geometrical configurations, 
it is customary\cite{ET:2025xjr} to plot Mollweide projection maps of the square root 
of the sum of squared detector pattern functions at a chosen time. 
In the present formalism, these correspond to the modulus of the corresponding 
\emph{detector vectors}.

\subsection{Ability to reconstruct the PMs from a pair of detectors}

To assess the ability of any detector pair to reconstruct the polarization modes, 
one begins by studying the correlation of the \emph{detector vectors} as a function of sky position. 
As explained previously, one expects to distinguish the polarization modes in 
sky regions where the correlation satisfies, for instance, 
$-0.95 \lesssim \rho \lesssim 0.95$. 
These limits are somewhat arbitrary and are suggested solely to indicate 
that in the complementary regions where $|\rho|>0.95$,
distinguishing between the two polarization modes becomes numerically more challenging.

For the purpose of reconstructing the polarization modes, 
it is also important to monitor the exterior product between the two 
\emph{detector vectors} of the corresponding observatories, since sky 
regions with very low values indicate probable numerical instability in 
the polarization mode reconstruction algebra. 

These new techniques can be applied to the study of efficiencies of
third-generation observatory proposals, and to the analysis of
GW detections; but this is out of the scope of the present work.

\subsection{The case of two observatories at different geolocations}

We have shown that the geometric concept of the \emph{plane of the GW event} 
is useful for discussing arrays of observatories sharing the same 
location and plane. 
This was made possible through the geometric concept 
of the \emph{detector vector}. 
However, it is important to recognize that these two concepts 
also apply to the case of two or more observatories 
at different locations and, of course, in different planes. 
More specifically, the notion of the \emph{plane of the GW event} 
is observatory-independent, it is an intrinsic property of the gravitational wave. 
Consequently, in this plane we can represent the \emph{detector vectors} for all observatories.

Of course, to construct the \emph{detector vectors},
one must have previously determined the sky location of the GW source, 
since the angular coordinates at each observatory are required. 
As we have shown, the polarization frame used to describe the GW does not affect the determination 
of the \emph{detector vectors}; however, choosing one is essential 
for the computational process.

Thus, constructing the \emph{detector vectors} for all detectors 
can be understood as a necessary intermediate step toward 
polarization mode reconstruction, 
since source sky localization is required for this task.

\section{Final comments}\label{sec:final}

By reformulating the discussion of GW source sky localization 
and polarization mode content, we have demonstrated the utility 
of the geometric concepts introduced here for analyzing transient GW events.

We have shown that this framework enables new discussions of GW detection properties 
by providing a novel perspective on recorded observations 
and several useful results.

From the definition of the \emph{detector vector}, 
it is clear that its norm quantifies the detector sensitivity 
for a particular GW event.

When two or more detectors are involved in the measurement of a GW, 
one can consider the sum of the \emph{detector vector} magnitudes, 
a quantity commonly used in previous studies. 
However, we also propose new diagnostic measures for detector pairs: 
the correlation $\rho_{X1,X2}$ between \emph{detector vectors} of observatories 
$X1$ and $X2$,
and the exterior product  $w_{X1,X2}$,
which is relevant for polarization mode reconstruction. 
We will apply these concepts in the analysis of 
GW observations and in sensitivity assessments for detector networks.

This new measure $\rho_{X1,X2}$ provides important information on the ability of 
a detector pair
to reconstruct the two PMs of the GW.
In particular it is clear that $\rho_{X1,X2}=1$ and $\rho_{X1,X2}=-1$
are cases where the reconstruction of two PMs is impossible;
since they correspond respectively to the case in which both detectors have recorded
the same PM component and the case in which the second detector has
recorded the negative of the first detector's component.

The set of results presented in section \ref{sec:geometry}
constitute a clear and useful framework for the discussion of several
topics involved in GW analysis.
For instance, theorem \ref{teo:2vec} is a convenient novel tool for the study
of co-located observatories.

This article has being dedicated to the introduction of these geometrical
constructions. 
In separate works, we will present detailed studies of expected sensitivities 
for different geometric designs of gravitational-wave observatories sharing the same site. 
In particular, we will discuss the complementarity of having 
a tristar detector in South America and a triangular detector in Europe.

We will also present in future work the application of 
these geometric tools to detected signals.

The framework presented in this article has the potential 
to clarify discussions of previous and future work related to GW source localization 
and polarization mode reconstruction.

\section*{Acknowledgments}

We thank Ezequiel Boero and Emanuel Gallo for a careful reading of the manuscript,
insightful comments
and suggesting several improvements.

We acknowledge the lifelong, standing support for our work from
Universidad Nacional de Córdoba(UNC) and CONICET.

\appendix

\section{Pattern functions}\label{app:pattern}

The pattern functions are\cite{Poisson2014}:
%{\footnotesize
\begin{align}\label{eq:patfun1}
F_+ =& \Big( \frac{1}{2}( 1 + \cos(\theta)^2) \cos(2 \phi) \cos(2 \psi)
- \cos(\theta) \sin(2 \phi) \sin(2 \psi) \Big) \sin(\chi) \\
F_\times =& \Big( \frac{1}{2}( 1 + \cos(\theta)^2) \cos(2 \phi) \sin(2 \psi)
+ \cos(\theta) \sin(2 \phi) \cos(2 \psi) \Big) \sin(\chi) \label{eq:patfun2} 
; 
\end{align}
where we are using the notation of the text book \cite{Poisson2014}
in which $(\theta,\phi)$ are the angular coordinates with respect to the
observatory orientation, $\psi$ denotes the third angular orientation to
the GW frame and $\chi$ is the angle between the arms of the observatory.

\section{Transformation properties of the PMs}\label{app:transfpm}

Let us recall that if $s_+$ and $s_\times$ denote the polarization components
with respect to an original basis, then the components with respect
to a basis rotated by an angle $\Delta\psi$ will be:
\begin{align}\label{eq:s+}
	s'_+ &= \cos(2 \Delta\psi) s_+ + \sin(2 \Delta\psi) s_\times \\
	\label{eq:sx}
	s'_\times &= -\sin(2 \Delta\psi) s_+ + \cos(2 \Delta\psi) s_\times 
	.
\end{align}

\section{Transformation properties of the pattern functions through $\Delta\psi$ rotations}\label{app:transfpattfunc}

Let us note that eqs. \ref{eq:patfun1} and \ref{eq:patfun2} can be expressed as
\begin{align}\label{eq:patfun1b}
	F_+ =&      f_1  \cos(2 \psi) - f_2 \sin(2 \psi)  \\
	F_\times =& f_1  \sin(2 \psi) + f_2 \cos(2 \psi) \label{eq:patfun2b} 
	; 
\end{align}
then an extra  $\Delta\psi$ rotation implies the following relations of the new components
in terms of the old ones
\begin{align}\label{eq:patfun1c}
	F'_+ =&      F_+  \cos(2 \Delta\psi) - F_\times \sin(2 \Delta\psi)  \\
	F'_\times =& F_+  \sin(2 \Delta\psi) + F_\times \cos(2 \Delta\psi) \label{eq:patfun2c} 
	;
\end{align}
which as expected is the inverse of the transformations \ref{eq:s+} , \ref{eq:sx}.

\section{Transformation properties of the pattern functions through $\Delta\phi$ rotations}\label{app:transfphi}

Note that the pattern functions can be expressed as
\begin{align}\label{eq:patfun3a}
	F_+ =&      f_a  \cos(2 \phi) - f_b \sin(2 \phi)  \\
	F_\times =& f_c  \sin(2 \phi) + f_d \cos(2 \phi) \label{eq:patfun3b} 
	; 
\end{align}
with $f_a = \frac{1}{2}( 1 + \cos(\theta)^2) \cos(2 \psi)  \sin(\chi)$,
$f_b = \cos(\theta) \sin(2 \psi) \sin(\chi)$, 
$f_c = \cos(\theta) \cos(2 \psi)  \sin(\chi)$
and
$f_d = \frac{1}{2}( 1 + \cos(\theta)^2) \sin(2 \psi) \sin(\chi)$.
With these, one can express the detector vector as
\begin{equation}\label{eq:s}
\vec{s}_X =
F_+ \hat s_+ + F_\times \hat s_\times
= \vec{s}_1 \cos(2 \phi) + \vec{s}_2 \sin(2 \phi)
,
\end{equation}
with
\begin{align}\label{eq:patfuns12a}
	\vec{s}_1  =&     ( f_a \hat s_+ + f_d \hat s_\times )   \\
	\vec{s}_2  =&  (f_c \hat s_\times - f_b \hat s_+ )    \label{eq:patfuns12b} 
	.
\end{align}

Note that
\begin{equation}
\begin{split}
< \vec{s}_1, \vec{s}_2 > =& - f_a f_b + f_c f_d \\
=&  -\frac{1}{2}( 1 + \cos(\theta)^2) \cos(2 \psi)  \cos(\theta) \sin(2 \psi) \sin^2(\chi) \\
&+\frac{1}{2}( 1 + \cos(\theta)^2)  \sin(2 \psi) \cos(\theta) \cos(2 \psi) \sin^2(\chi) \\
=& \, 0
,
\end{split}
\end{equation}
and that
\begin{equation}
\begin{split}
	< \vec{s}_1, \vec{s}_1 > =&  f_a^2 + f_d^2 \\
	=&  \Big(\frac{1}{2}( 1 + \cos(\theta)^2) \cos(2 \psi)  \sin(\chi) \Big)^2 \\
  	&+  \Big(\frac{1}{2}( 1 + \cos(\theta)^2) \sin(2 \psi)  \sin(\chi) \Big)^2 \\
	=& \Big(\frac{1}{2}( 1 + \cos(\theta)^2)  \sin(\chi) \Big)^2
	,
\end{split}
\end{equation}
and also
\begin{equation}
\begin{split}
	< \vec{s}_2, \vec{s}_2 > =&  f_c^2 + f_b^2 \\
	=&  \Big(\cos(\theta) \sin(2 \psi)  \sin(\chi) \Big)^2 \\
	&+  \Big(\cos(\theta) \cos(2 \psi)  \sin(\chi) \Big)^2 \\
	=& \Big( \cos(\theta)  \sin(\chi) \Big)^2
	.
\end{split}
\end{equation}
This means that $\{\vec{s}_1, \vec{s}_2\}$ are an orthogonal basis, with in general
unequal magnitudes. 
Therefore as a function of $\phi$ the detector vector of eq. \ref{eq:s}
changes components with cosine and sine functions with argument $2\phi$, marking the points of an ellipse.
In turn, continuing the discussion above, for two detectors
at the same site with an angle $\Delta \phi$,
the corresponding  vector, for the second observatory, in the signal plane,
will have argument $2(\phi - \Delta \phi)$ for the cosine and sine components.

More concretely, let
\begin{equation}\label{eq:s1}
	\vec{s}_{X1} =
	F_+ \hat s_+ + F_\times \hat s_\times
	= \vec{s}_1 \cos(2 \phi) + \vec{s}_2 \sin(2 \phi)
	,
\end{equation}
be the \emph{detector vector} for observatory 1; where the dependence on the coordinate
$\phi$ is shown.
Then let the observatory 2 be at the same place but rotated an  angle $\Delta \phi$
with respect to the first.
Then, at observatory 2, using coordinate $\phi'$; they have
$\phi'=\phi - \Delta \phi$, so that one can write
\begin{equation}\label{eq:s2}
	\vec{s}_{X2} =
	F_+ \hat s_+ + F_\times \hat s_\times
	= \vec{s}_1 \cos(2 \phi - 2 \Delta \phi) + \vec{s}_2 \sin(2 \phi - 2 \Delta \phi)
	.
\end{equation}
Note that
\begin{equation}\label{eq:s1dots2}
\begin{split}
< \vec{s}_{X1} ,	\vec{s}_{X2}> =&
< \vec{s}_1, \vec{s}_1 > \cos(2 \phi) \cos(2 \phi - 2 \Delta \phi) \\
&+	< \vec{s}_2, \vec{s}_2 > \sin(2 \phi)\sin(2 \phi - 2 \Delta \phi) \\
=&
\Big(\frac{1}{2}( 1 + \cos(\theta)^2)  \sin(\chi) \Big)^2
\cos(2 \phi) \cos(2 \phi - 2 \Delta \phi) \\
&+
\Big( \cos(\theta)  \sin(\chi) \Big)^2
\sin(2 \phi)\sin(2 \phi - 2 \Delta \phi)
;
\end{split}
\end{equation}
so that for $\Delta \phi = \pm \frac{\pi}{4}$ one has
\begin{equation}\label{eq:s1dots2_pis2}
\begin{split}
< \vec{s}_{X1} ,	\vec{s}_{X2}> =&
\Big(\frac{1}{2}( 1 + \cos(\theta)^2)  \sin(\chi) \Big)^2
\cos(2 \phi) \cos(2 \phi \mp \frac{\pi}{2}) \\
&+
\Big( \cos(\theta)  \sin(\chi) \Big)^2
\sin(2 \phi)\sin(2 \phi \mp \frac{\pi}{2}) \\
=&
\Big(\frac{1}{2}( 1 + \cos(\theta)^2)  \sin(\chi) \Big)^2
\cos(2 \phi) (\pm) \sin(2 \phi) \\
&+
\Big( \cos(\theta)  \sin(\chi) \Big)^2
\sin(2 \phi) (\mp) \cos(2 \phi) \\
=&
\pm \cos(2 \phi) \sin(2 \phi)  \sin(\chi)^2
\Bigg(
\frac{1}{4}	-\frac{1}{2}\cos(\theta)^2 + \frac{1}{4}\cos(\theta)^4
\Bigg) \\
=&
\pm \frac{1}{2} \sin(4 \phi)  \sin(\chi)^2
\Bigg(
\frac{1}{4}	-\frac{1}{2}\cos(\theta)^2 + \frac{1}{4}\cos(\theta)^4
\Bigg) \\
=&
\pm \frac{1}{8} \sin(4 \phi)  \sin(\chi)^2 \sin(\theta)^4
.
\end{split}
\end{equation}

\section{Triangular configuration null stream}\label{app:triangular}

We have seen previously in theorem \ref{teo:2vec} that the three vectors
points to different positions on an ellipse.
Then let say the first vector is at
\begin{equation}\label{eq:vx1}
	\vec{s}_{X1} = \cos(2 \phi) \vec{s}_1 + \sin(2 \phi) \vec{s}_2
	.
\end{equation}
Then one has for the second vector
\begin{equation}\label{eq:vx2}
	\vec{s}_{X2} = \cos(2 \phi + \frac{2\pi}{3}) \vec{s}_1 + \sin(2 \phi + \frac{2\pi}{3}) \vec{s}_2
	;
\end{equation}
while for the third vector one has
\begin{equation}\label{eq:vx3}
	\vec{s}_{X3} = \cos(2 \phi - \frac{2\pi}{3}) \vec{s}_1 + \sin(2 \phi - \frac{2\pi}{3}) \vec{s}_2
	.
\end{equation}
Let us note that
\begin{equation}\label{eq:coef_s1}
	\cos(2 \phi)+\cos(2 \phi + \frac{2\pi}{3})+\cos(2 \phi - \frac{2\pi}{3})
	=
	\cos(2 \phi)+2 \cos(2 \phi) \cos(\frac{2\pi}{3})) = 0
	;
\end{equation}
and that
\begin{equation}\label{eq:coef_s2}
	\sin(2 \phi)+\sin(2 \phi + \frac{2\pi}{3})+\sin(2 \phi - \frac{2\pi}{3})
	=
	\sin(2 \phi)+2 \sin(2 \phi) \cos(\frac{2\pi}{3})) = 0
	.
\end{equation}
That is one has
\begin{equation}\label{eq:sumxvecs}
	\vec{s}_{X1} + \vec{s}_{X2} +\vec{s}_{X3} = 0
	.
\end{equation}

Note that the algebra also goes through if one defines
\begin{align}\label{eq:patfuns12c}
	{s}_1  =&     ( f_a  s_+ + f_d  s_\times )   \\
	{s}_2  =&  (f_c  s_\times - f_b  s_+ )    \label{eq:patfuns12d} 
	;
\end{align}
that is, one also has
\begin{equation}\tag{\ref{eq:sumxsig}}
	s_{X1} + s_{X2} + s_{X3} = 0
	;
\end{equation}
which is sometimes called the null stream.

%%%%%%%%%%%%%%%%%%%%%%%%%%%%%%%%%%%%%%%%%%%%%%%%%%%%%%%%%%%%%%%%%%%%%%%%%%%%%%%%%%
%\bibliography{/home/moreschi/biblio/refosv}
%%%
%%%
%\bibliographystyle{/home/moreschi/biblio/bibstyles/utphys} %%%%

%\bibliographystyle{abbrv}
%\bibliographystyle{plain}
%\bibliographystyle{numbers}
%\bibliographystyle{/home/moreschi/biblio/bibstyles/unsrt}

%\bibliographystyle{/home/moreschi/biblio/bibstyles/utphys-2-7} %%%%

%\bibliographystyle{/home/moreschi/biblio/bibstyles/kp-osvaldo} %%%%%%%%%%%
%\bibliographystyle{abbrv}
%%
%%%\bibliographystyle{alpha}
%%%\bibliographystyle{/home/moreschi/biblio/bibstyles/halpha}
%%%\bibliographystyle{/home/moreschi/biblio/bibstyles/utcaps}
%%%\bibliographystyle{/home/moreschi/biblio/bibstyles/h-physrev}
%%%\bibliographystyle{/home/moreschi/biblio/bibstyles/hapalike}
%%%\bibliographystyle{/home/moreschi/biblio/bibstyles/hieeetr}
%%%\bibliographystyle{/home/moreschi/biblio/bibstyles/hsiam}
%%%\bibliographystyle{/home/moreschi/biblio/bibstyles/hunsrt}
%%%\bibliographystyle{/home/moreschi/biblio/bibstyles/h-elsevier}
%%
%\bibliographystyle{/home/moreschi/biblio/bibstyles/kp}
%%

\end{document}